\PassOptionsToPackage{table}{xcolor}
\documentclass[sigconf]{acmart}
\usepackage{hhline}
\usepackage{pifont}
\AtBeginDocument{%
  \providecommand\BibTeX{{%
    \normalfont B\kern-0.5em{\scshape i\kern-0.25em b}\kern-0.8em\TeX}}}

\copyrightyear{2020}
\acmYear{2020}
\setcopyright{acmcopyright}
\acmConference[FDG '20]{International Conference on the Foundations of Digital Games}{September 15--18, 2020}{Bugibba, Malta}
\acmBooktitle{International Conference on the Foundations of Digital Games (FDG '20), September 15--18, 2020, Bugibba, Malta}
\acmPrice{15.00}
\acmDOI{10.1145/1122445.1122456}
\acmISBN{978-1-4503-8807-8/20/09}

\begin{document}

\title{Exploring Help Facilities in Game-Making Software}

\author{Dominic Kao}
\affiliation{%
  \institution{Purdue University}
  \streetaddress{610 Purdue Mall}
  \city{West Lafayette}
  \state{Indiana}
  \postcode{47906}
}
\email{kaod@purdue.edu}


\begin{abstract}
	Help facilities have been crucial in helping users learn about software for decades. But despite widespread prevalence of game engines and  game editors that ship with many of today's most popular games, there is a lack of empirical evidence on how help facilities impact game-making. For instance, certain types of help facilities may help users more than others. To better understand help facilities, we created game-making software that allowed us to systematically vary the type of help available. We then ran a study of 1646 participants that compared six help facility conditions: 1) Text Help, 2) Interactive Help, 3) Intelligent Agent Help, 4) Video Help, 5) All Help, and 6) No Help. Each participant created their own first-person shooter game level using our game-making software with a randomly assigned help facility condition. Results indicate that Interactive Help has a greater positive impact on time spent, controls learnability, learning motivation, total editor activity, and game level quality. Video Help is a close second across these same measures. 
\end{abstract}




\maketitle

\section{Introduction}
Many successful video games, such as \textit{Dota 2} and \textit{League of Legends} (from \textit{WarCraft 3}), \textit{Counter-Strike} (from \textit{Half-Life}), and the recent \textit{Dota Auto Chess} (from \textit{Dota 2}), are modifications of popular games using game-making or level-editing software. The popular game engine Unity powers 50\% of mobile games, and 60\% of all virtual reality and augmented reality content \cite{Unity}. Despite the reach and impact of game-making, very few empirical studies have been done on help facilities in game-making software. For example, in our systematic review of 85 game-making software, we find that the majority of game-making software incorporates text help, while about half contain video help, and only a small number contain interactive help. Given the large discrepancies in help facility implementation across different game-making software, it becomes important to question if different help facilities make a difference in user experience, behavior, and the game produced.

Help facilities can teach users how to use game-making software, leading to increased quality in created games. Through fostering knowledge about game-making, help facilities can better help novice game-makers transition to becoming professionals. While studies on game-making and help facilities do not currently exist, there is good motivation for this topic from gaming. A key study by Andersen et al. \cite{Cooper2012} suggests that help facilities can be beneficial in complex games (increasing play time by as much as 29\%), but their effects were non-significant in simpler games where mechanics can be discovered through experimentation. Because game-making software often presents users with a larger number and higher complexity of choices compared to  games \cite{Hayes2008a}, game-making is likely a domain in which help facilities play an important role.

In this paper, we start by first reviewing the help facilities in popular game-making software, including game engines and game editors. This allowed us to understand which types of help facilities are present in game-making software, as well as how those help facilities are implemented. This review directly influenced the design of our help facility conditions in our main study. We then describe our game-making software, \textit{GameWorld}, which allows users to create their own first-person shooter (FPS) games. Lastly, we describe a between-subjects experiment conducted on Amazon Mechanical Turk that varied the help facility available to the user. This allowed us to isolate the impact of help facility type while keeping all other aspects of the game-making software identical. In this experiment, we had 5 research questions:

\noindent\textbf{RQ1:} Do help facilities lead to higher motivated behavior (time spent, etc.)? \\
\textbf{RQ2:} Do help facilities improve learnability of controls? \\
\textbf{RQ3:} Do help facilities improve learning motivation? \\
\textbf{RQ4:} Do help facilities improve cognitive load? \\
\textbf{RQ5:} Do help facilities improve created game levels? \\
\textbf{RQ6:} Does time spent on help facilities vary?

Results show that the interactive help has a substantial positive impact on time spent, controls learnability, learning motivation, cognitive load, game-making actions, and final game level quality. The video help has a similarly positive impact on time spent, learning motivation, cognitive load, and final game level quality.

On the other hand, results show that having no help facility results in the least amount of time spent, lowest controls learnability, lowest learning motivation, highest cognitive load, lowest game-making actions, and lowest final game level quality. We found that the other help facility conditions (text, intelligent agent, all) generally did not significantly differ from no help, except in cognitive load (text is better than no help, but worse than all other conditions). Finally, we conclude with a discussion of design implications based on the results of the study.

\section{Related Work}

HCI and games researchers have long been interested in games and learning \cite{Gutwin2016,Iacovides2014,Johanson2016a,Wauck2017,VisaniScozzi2017,Rojas2013,Harpstead2013,Harteveld,Kao2016f,Kao2015a,Kao2019}. This includes studies on tutorials \cite{Cooper2012}, differences in frequent versus infrequent gamers' reactions to tutorials \cite{Moirn2016}, leveraging reward systems \cite{Gaston2017}, encouraging a growth mindset, or the idea that intelligence is malleable \cite{Rourke2016,ORourke2014,kaoexploring}, avatars \cite{Kao2018,Kao2016e,Kao2015,Kao2019a}, embellishment \cite{Kao2017,Kao2020}, and many more. AI and games researchers have also begun to take interest, such as the automatic generation of video game tutorials  \cite{Green2018,Green2018a}, and the adaptation of tutorials to individual user skill levels \cite{Aytemiz2018}.%

In this section, we begin with an overview of software learnability and multimedia learning. We then review the types of help facilities that are found most often in game-making software. These are text documentation, video tutorials, and interactive tutorials. We also investigate intelligent agents. Despite not being present in most game-making software, intelligent agents have been widely explored as an effective means of teaching in the academic literature. Although there are many other types of help (e.g., showing a random tip on startup) and variations thereof \cite{Cooper2012}, our focus is on: 1) Core help facilities commonly available in game-making software, and 2) Common implementations of those facilities. Both this literature, and the review of game-making software, provides a baseline for developing our own help facility conditions.

\subsection{Software Learnability}

Software learnability is a general term that refers to learning how to use a piece of software. Software learnability can be measured along different dimensions, including task metrics (i.e., task performance), command metrics (i.e., based on commands issued by the user), mental metrics (i.e., related to cognitive processes), and subjective metrics (i.e., learnability questionnaires). In this paper, we triangulate across these multiple categories by leveraging expert game level ratings (task), total game-making actions (command), cognitive load measures (mental), and a questionnaire assessing learnability of controls (subjective), to gauge the effects of help facilities on game-making software learnability. One important aspect of our study is cognitive load---this refers to human working memory usage \cite{Sweller1988}. Here, we are interested in studying the amount of cognitive load experienced by users in each of the help facility conditions. Although help facilities may help users moderate cognitive load through scaffolding the game-making activity, they may also lead to negative impacts, e.g., overwhelming the user with information \cite{Wise2009}.

\subsection{Multimedia Learning Theory}

Multimedia learning theory illustrates the principles which lead to the most effective multimedia (i.e.,  visual and auditory) teaching materials \cite{mayer2002multimedia}. These principles include the \textit{multimedia principle} (people learn better from words and pictures than from words alone), the \textit{spatial contiguity principle} (people learn better when corresponding words and pictures are presented near rather than far from each other), and the \textit{temporal contiguity principle} (people learn better when corresponding words and pictures are presented simultaneously rather than successively) \cite{mayer2006ten}. We utilize this framework as one internal guide in developing our help facilities. In the remaining sections, we survey different modalities of help facilities.

\subsection{Text-Based Help Facilities}

Early forms of computing documentation originated from the advent of commercial mainframe computing \cite{zachry2001constructing}. Much of the research on text help is dedicated to improving the user experience of computing documentation. Converging evidence suggests that user frustration with computers is a persistent issue that has not been satisfactorily ameliorated by accompanying documentation \cite{lazar2006workplace,Mirel1998,Barnsley2013}. Over the past few decades, researchers have proposed several methods for improving the user experience of computing documentation, including standardizing key software terminology and modes of expression \cite{Antoniol2003,Wang2009}; automatically generating documentation material \cite{Paris2007}; using semantic wiki systems to improve and accelerate the process of knowledge retrieval \cite{DeGraaf2011}; and drastically shortening text manuals by eliminating large sections of explanation and elaborations \cite{Carroll1987}. A significant issue in this research, however, is the dearth of systematic reviews and comprehensive models for evaluating the efficacy of text-based help facilities \cite{Zhi2015}. As a result, it remains difficult to determine both the utility of computing documentation for users and developers and whether the benefits of production outweigh the costs \cite{DeSouza2005}.

In one study of tutorials and games, text-based tutorials were associated with a 29\% increase in length of play in the most complex game; there was no significant increase with the tutorials for simpler games, which suggests that investing in the creation of tutorials for simpler games may not be worth it \cite{Cooper2012}. Researchers have stated that official gaming documentation faces a gradual but substantial decline \cite{greene2011s}. This can be attributed to several factors. Scholars and consumers of computer games typically agree that the best gaming experience is immersive and, therefore, that necessitating any documentation to understand gameplay is a hindrance; at the same time, complex games that lack text-based help facilities are frequently criticized for having steep learning curves that make immersion difficult \cite{moeller2016computer}. Moreover, researchers have argued that there is a lack of standardization across different games, and that help documentation is often written by game developers themselves (often through simply augmenting internal development texts), which has decreased the efficacy of text-based documentation \cite{Winget2011,moeller2016computer,greene2011s,ambler2012best}.

\subsection{Interactive Tutorial Help Facilities}

Since the mid-1980s, early research has sought to understand the effectiveness of interactive tutorials on learning \cite{Charney1986,Lieberman2002,Lieu1999}. Interactive tutorials have been found to be especially effective in subjects that benefit from visualizing concepts in detail; engineering students, for example, can interact with graphical representations of objects that are difficult or impossible to do so with still images \cite{Lieu1999}. Interactive tutorials have been found to be highly effective in learning problem-solving \cite{Frith2004,tiemann1990effects}. Additionally, interactive tutorials have been found to be superior to non-interactive methods in learning factual knowledge \cite{Lieberman2002}, database programming \cite{Pahl2002}, medical library instruction \cite{Anderson2009}, and basic research skills \cite{stiwinter2013using}.

Game designers often emphasize the importance of user experimentation while learning new concepts \cite{ray2010tutorials}. This experimentation, James Gee argues, should take place in a safe and simplified environment where mistakes are not punished \cite{Gee2005}. Kelleher and Pausch have shown that restricting user freedom improves tutorial performance \cite{Kelleher2005}. Using the Alice programming environment, they find that with an interactive tutorial called Stencils, users are able to complete the tutorial faster and with fewer errors than a paper-based version of the same tutorial. Molnar and Kostkova found that children 10-13 years of age reacted positively to the incorporation of an interactive tutorial that guides the player explicitly through game mechanics \cite{Molnar2013}. On the other hand, participants that did not play the interactive tutorial found the game more awkward \cite{Molnar2014}. Frommel et al. found that in a VR game, players taught more interactively had higher positive emotions and higher motivation \cite{Frommel2017}.

\begin{figure*}[!ht] 
\begin{minipage}[t]{0.32\linewidth}
\centering
\includegraphics[width=1\linewidth]{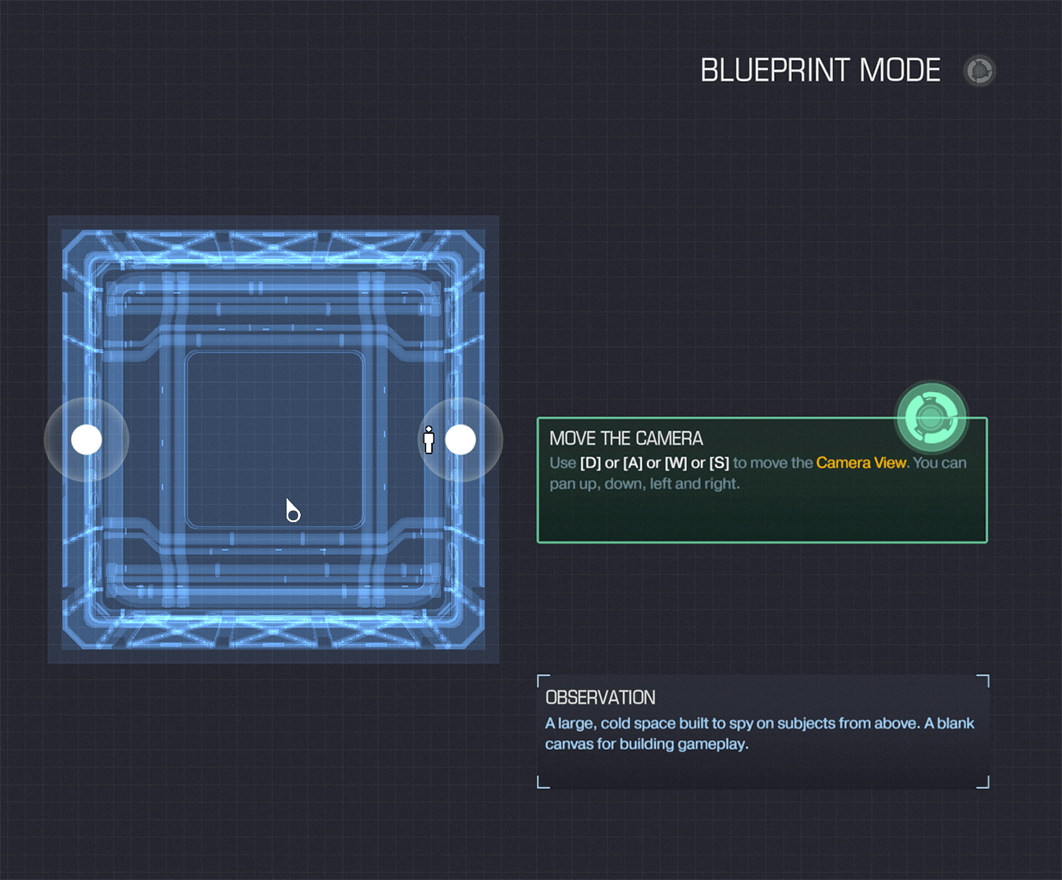}
\caption{Doom's SnapMap interactive tutorial.}
\Description{Fig1 description}
\label{fig:EditorExampleDoom}
\end{minipage}%
\hfill
\begin{minipage}[t]{0.32\linewidth}
\centering
\includegraphics[width=1\linewidth]{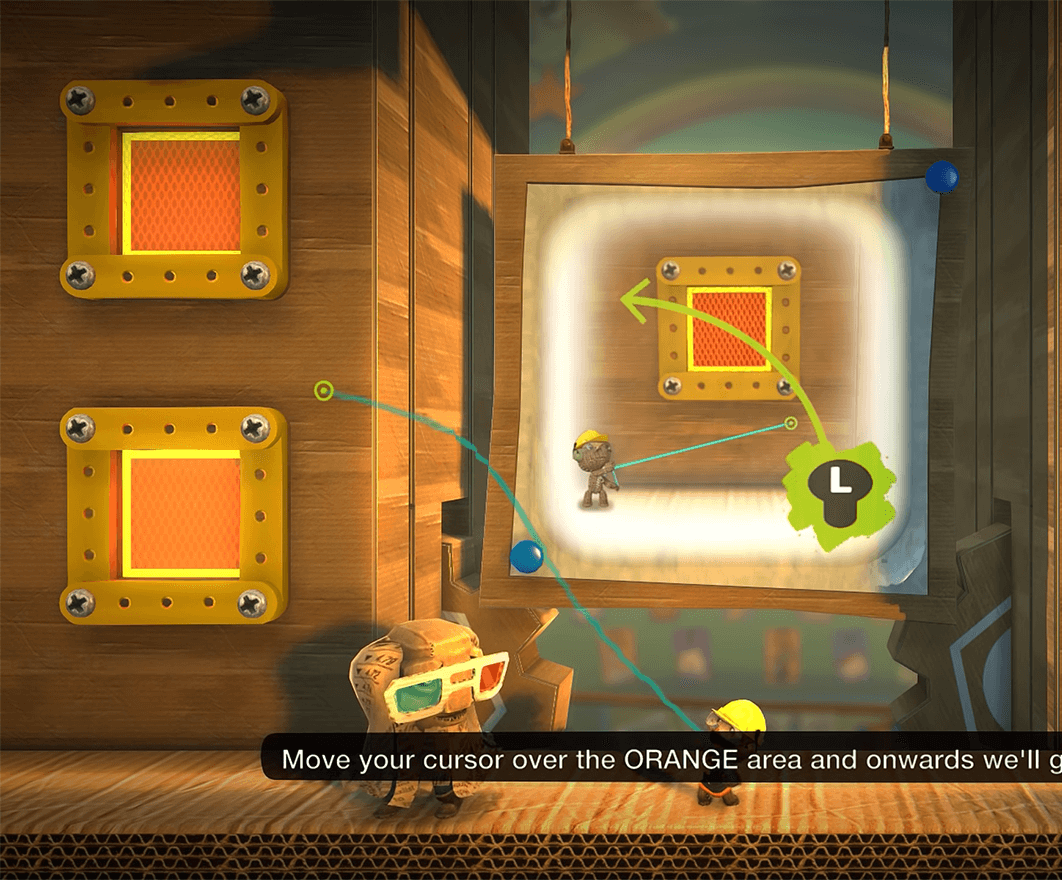}
\caption{LittleBigPlanet 3 interactive tutorial.}
\Description{Fig2 description}
\label{fig:EditorExampleLBP3}
\end{minipage}%
\hfill
\begin{minipage}[t]{0.32\linewidth}
\centering
\includegraphics[width=1\linewidth]{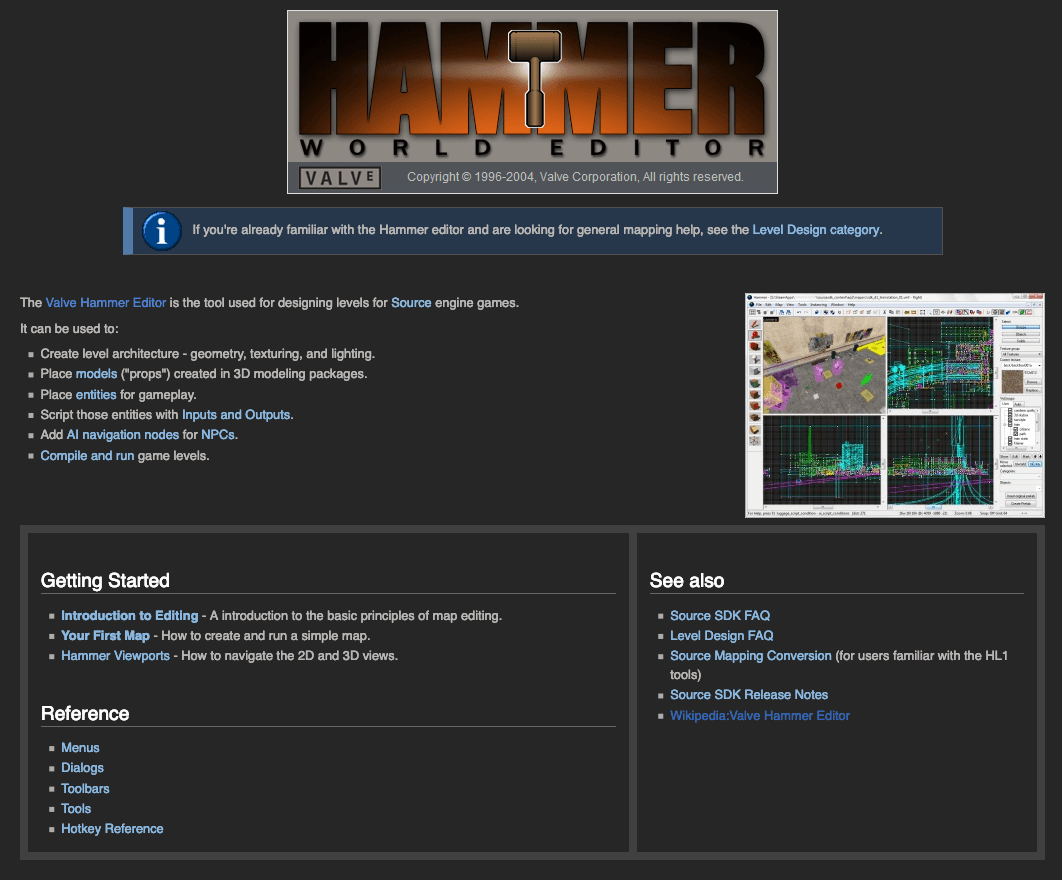}
\caption{Valve's Hammer editor text doc.}
\Description{Fig3 description}
\label{fig:EditorExampleHammer}
\end{minipage} 
\end{figure*}

\subsection{Intelligent Agent Help Facilities}

The persona effect was one of the earliest studies that revealed that the mere presence of a life-like character in a learning environment increased positive attitudes \cite{Lester1997,Kao2016b}. Intelligent agents are on-screen characters that respond to feedback from users in order to enhance their experience \cite{Pena2002,Wooldridge1995}. These are often used to effectively tailor learning environments for individual students \cite{Pena2002,Wooldridge1995}. Intelligent agents can help to personalize learning more effectively than standard teaching tools \cite{Pena2002}, and allow for  human-like interaction between the software and the user that would not otherwise be possible \cite{Selker2002,baylor1999intelligent}. Intelligent agents have been integrated into several games whose purpose is to teach the player. The TARDIS framework uses intelligent agents in a serious game for social coaching for job interviews \cite{Anderson2013a}. Other educational games have utilized intelligent agents to teach the player number factorization \cite{Conati2004,Conati2002}, the Java compilation process \cite{Gomez-martin}, and computational algorithms \cite{Gamage2018}. %

\subsection{Video Tutorial Help Facilities}

Video-based tutorials utilize the modalities of audio, animation, and alphabetic text. Research has shown that user performance is increased when animation is combined with an additional semiotic mode, such as sound or words \cite{Mayer1991}. Video animation is effective for recall when illustrating highly visual facts, concepts, or principles \cite{rieber1994computers} (p.116). For instance, video tutorials can display a task sequence in the same way a user would see it on their own computer screen, leading to congruence between the video and the real-life task execution \cite{Tversky2002}. Many studies have shown that video tutorials can be highly effective \cite{Brecht2008,van2014comparison,xeroulis2007teaching}. For example, one study found that 24.2\% of students without videos failed a course on introductory financial accounting, whereas the failure rate was only 6.8\% among students that had the videos available \cite{Brecht2008}. Another study that compared text tutorials to video tutorials for learning software tools found that both types of tutorials had their advantages. Namely, video tutorials were preferred for learning new content; however, text tutorials were useful for looking up specific information \cite{Kafer2017}. Video walkthroughs are common instructional tutorials used in games to help players overcome a game's challenges through the imitation of actions \cite{Carr2005,nylund2015walkthrough,Bergstrom2018}. For example, a classroom study supplemented the use of video games with walkthroughs, and found that students found the video-based walkthroughs more helpful than the text-based ones \cite{Bergstrom2018}.

\subsection{Game-Making}

Academic interest in game-making has its roots in constructionism: the theory of learning in which learners construct mental models for understanding the world \cite{Papert1991}. Early manifestations  included ``Turtle Geometry,'' an environment for programming an icon of a turtle trailing lines across a computer display. Research at the intersection of HCI, game-making, and education has shown that game-making has promise for increasing engagement, knowledge, and skills in a variety of domains \cite{Kafai2006,Kafai2015b,Hayes2008,RobertsonJudy.Robertson@hw.ac.uk2013,Robertson2012,Robertson2008,Denner2012,Allsop2016,Howland2015,Gee,Kao2017b}. However, despite an extensive literature on game-making and education, game-making \textit{software} is seldom studied. \cite{Malbos2013} is one rare example in which 8 game-making tools were contrasted on their immersive features. Given the scarcity of work on game-making software, it is difficult to predict which types of help facilities will be most effective. Even in games, despite employing a wide variety of tutorial styles, the relative effectiveness of these styles is not well understood \cite{Cooper2012}. The main goal of the current study is to explore the effects of help facilities within game-making software.

\section{Intersectionality Between Play and Making}

Before proceeding, it is crucial to discuss our approach in studying game-making software. Frequently, in developing this work and discussing it with others, we often broached the topic of what \textit{making} is, and what \textit{play} is. Yet in trying to define these terms, even in a specific context such as games, we reach a deep philosophical impasse. Huizinga is well-known to be the progenitor of one of the most widely used (but widely contested) definitions of \textit{play} \cite{huizinga2014homo,Salen2004}. Piaget made the following observation: ``the many theories of play expounded in the past are clear proof that the phenomenon is difficult to understand'' \cite{Piaget2013}. Instead of attempting to delineate the two terms, we argue that it is precisely their intersectionality that needs further theoretical and empirical grounding. We argue that strictly categorizing an activity as \textit{play} or \textit{making} threatens to constrain researchers to drawing on traditional epistemologies inherent to how the terms have been defined previously, rather than building new interdisciplinary bridges which shed light on both parallels and divergences.

For example, we find in the next section that game-making software often appears to fall on a continuum that is neither fully software nor fully game. We could argue that LittleBigPlanet should be categorized as a game, and that Unity should be categorized as software. Yet elements of play and making are present even in these more extreme examples---in Unity, users engage in frequent play-testing, in part to see if their created game is ``fun''. Therefore, there appear to be a number of both parallels and divergences between play and making, and their degree of overlap in any given context will inevitably depend on the definitions that one has decided to apply. In this paper, we avoid strict categorization of game-making software as being a pure ``game'' or pure ``software''---this allows our survey to more flexibly encompass a wide range of game-making systems, regardless of whether they exist as independent environments or embedded inside the ecology of a game.

\section{Review of Game-Making Software}

Before designing our experiment, we reviewed 85 different game-making software. This includes both game engines and official level editors. We retrieved the list of software based on commercial success and popularity (Wikipedia/Google), critical reception (Metacritic), and user reception (Slant.co). For example, Slant.co shows user-driven rankings for ``What are the best 2D game engines?'' and ``What are the best 3D game engines?''.

Each piece of software was first installed on an appropriate device, then explored by 2 experienced (8+ years of professional game development experience) game developers independently for 1 hour. Each individual then provided their own summary of the software and the help facilities available. At this stage, all possible help facilities were included, such as interactive tutorials, startup tips, community forums, and so on. In some rare instances, we excluded software prior to review that did not come from an official source. (One notable example is Grand Theft Auto V, which does not have official modding tools.) For examples, see Figure~\ref{fig:EditorExampleDoom},~\ref{fig:EditorExampleLBP3}, and~\ref{fig:EditorExampleHammer}.

Next, we condensed our review into a table summarizing the different types of help available for each software. The table was coded independently by the 2 developers, then discussed and re-coded repeatedly until consensus was reached. At this stage, we made the distinction between help facilities contained directly in the software versus external help facilities. External help facilities included online courses, community forums, and e-mailing support. These types of help fall outside the main intent of our current research, which is to study help facilities contained in the software itself and were therefore excluded. An exception was made for common internal help facilities that were external, so long as they came from an official source and so long as a link was included to those help facilities directly from within the software (e.g., online text documentation, videos, etc.). Unofficial sources of help were not included. Finally, types of help that were contained within the software but were not substantive enough to warrant their inclusion as a core help facility (such as a random tip appearing each time the software boots) were excluded.

\begin{figure}[ht!]
 \centering
\includegraphics[width=230pt]{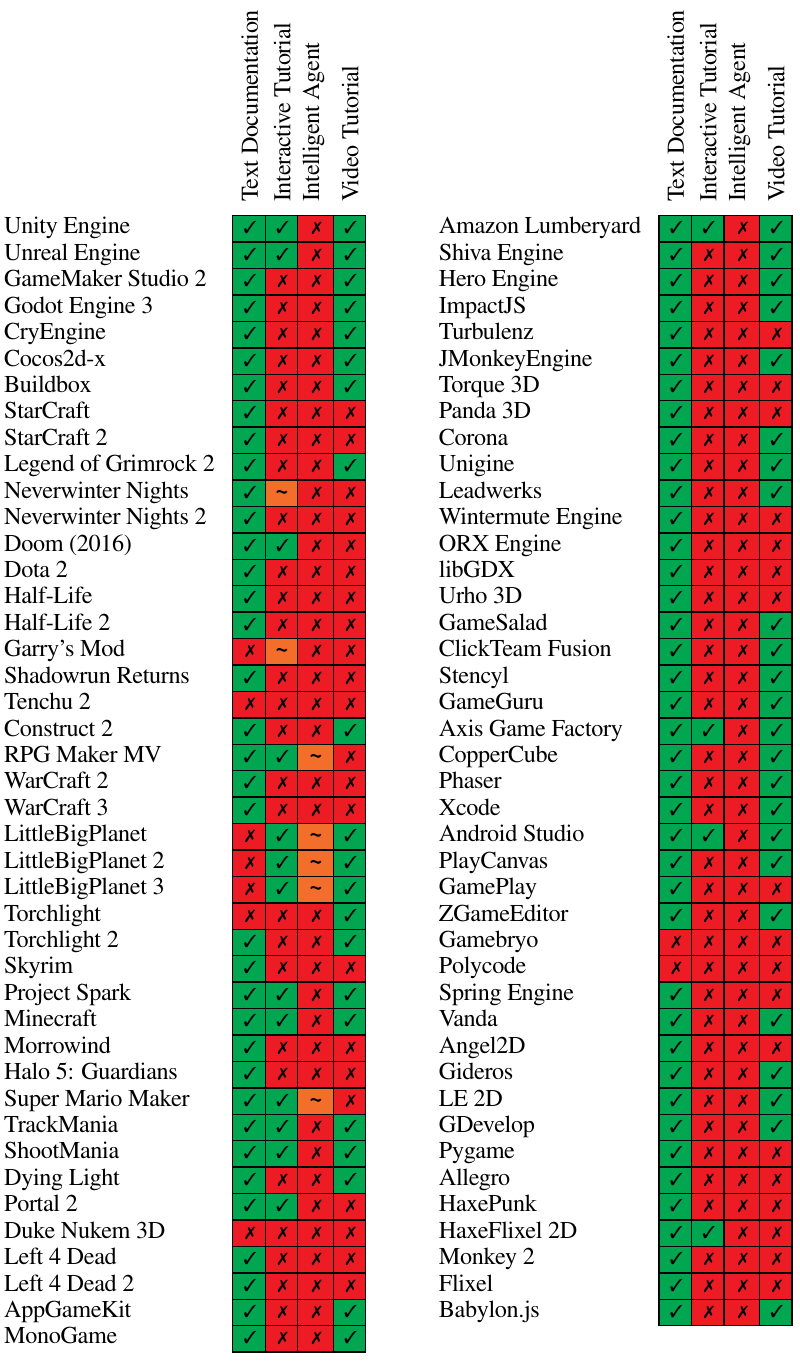} 
\caption{Game-making software and their official help facilities. Green means ``Yes'' (\checkmark), orange means ``Somewhat'' ($\mathbf{\sim}$), and red means ``No'' ($\mathbf{\times}$).}
\Description{Fig4 description}
\label{fig:SystemsReview}
\end{figure}

Our final table contained the following categories of help facilities: text documentation, interactive tutorial, and video tutorial. In addition, intelligent agent was included as a result of our earlier rationale. See Figure \ref{fig:SystemsReview}. Overall, the review shows that text documentation is prevalent in the majority of game-making software (89.4\%). Official video tutorials are present in approximately half of game-making software (52.9\%). A lesser number of game-making software contain interactive tutorials (20.0\%). Finally, no games contained an intelligent agent that responded to user choices (0.0\%). A few game-making software contained a character that would lead the player through a tutorial, but these were purely aesthetic and were not full-fledged intelligent agents.

\begin{figure*}[ht!]
\centering
\includegraphics[width=\linewidth]{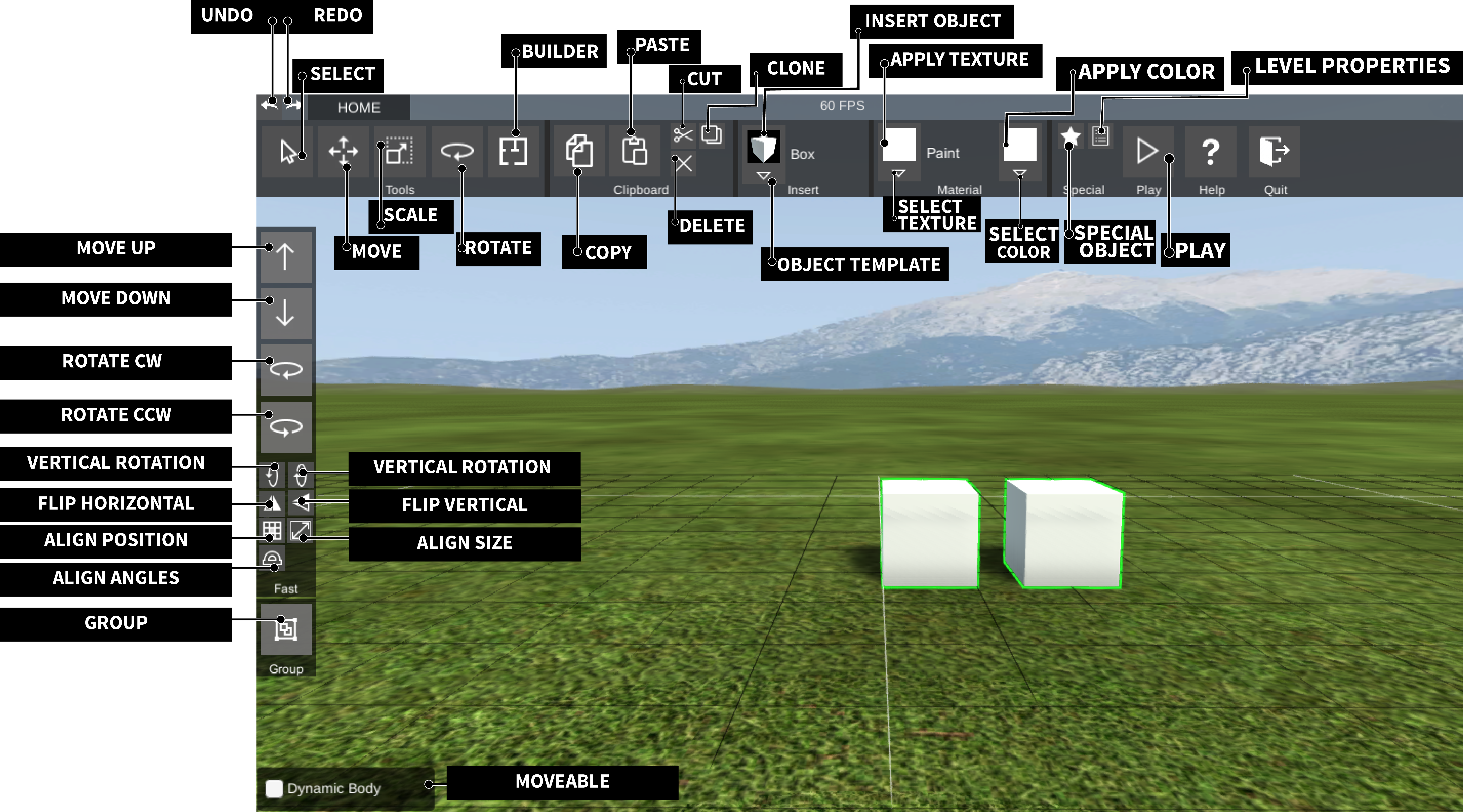}
\caption{Interface overview. Each interface element has a corresponding tooltip.}
\Description{Fig5 description}
\label{fig:InterfaceOverview}
\end{figure*}

\section{The Game-Making Software}

We developed a game-making software called \textit{GameWorld}\footnote{Demo: \url{https://youtu.be/O7_VH0IyWdo}}. \textit{GameWorld} was developed using a spiral HCI approach by repeatedly designing, implementing, and evaluating prototypes in increasingly complex iterations. Evaluation of prototypes was performed with experienced game developers known to the author. \textit{GameWorld} was developed specifically for novice game-makers, and allows users to create a first-person shooter game without any coding.

Figure \ref{fig:InterfaceOverview} shows the main interface elements. The top of the interface is primarily dedicated to object-related actions. The left side allows additional object manipulations. For example, objects in \textit{GameWorld} are typically aligned to an underlying grid. However, the user can hold down Control while modifying position, rotation, or scale, which ignores the grid alignment and gives the user more flexibility. Therefore, the ``Align'' buttons allow for objects to be snapped back into grid alignment. Objects can also be grouped (for easier management), and be made dynamic (which means they are moveable during play, for instance from collisions with bullets or player models). Dynamic is an important modifier for certain objects, such as a door, which consists of a door frame, a door joint, and a door which has the dynamic modifier enabled.

\textit{Objects.} There are 36 pre-made objects that users can place. These include simple objects (e.g., a sphere), to more complex objects (e.g., a guard room). Players can also create their own objects, for example by grouping objects together and saving them as a ``pre-fab''. Objects can be textured and colored. There are 76 pre-made textures that users can choose. There are 146 color choices. %

\textit{Special Objects.} Special objects are non-standard objects like door joints, invisible walls, lights, player and enemy spawn points, and trees. These are manipulated in the same way as normal objects.

\textit{Level Properties.} Within the level properties menu, players can modify the starting health of the player, number of enemies, whether enemies should respawn after death (and how often), certain modes useful for testing (e.g., player invincibility), etc. Some of these settings can also be changed during play testing in the pause menu.%

\textit{Builder Tool.} The builder tool allows users to create arbitrary objects using cubes, each cube corresponding to one grid volume.

\section{Developing Help Facilities}

In developing the help facility conditions, our primary objectives were: 1) Consistent quality across the different help facilities, and 2) Realistic implementations similar to current game-making software. To this end, we sought freelancer game developers to help with ``Providing Feedback on Game-Making Software''. We told game developers that we wanted critical feedback on game-making software being developed. We hired a total of 15 professional game developers, each with an average of 4 years (SD=2.0) of game development experience. Each game developer had worked with at least 3 different game engines, with more than half of the developers having experience with 5+. Developers all had work experience and portfolios which reflected recent game development experience (all within one year). These game developers provided input throughout the help facility development process. Game developers provided feedback at three different times during the the creation of our help facility conditions: During the initial design, after initial prototypes, and after completing the polished version. Game developer feedback was utilized to create help facilities that developers thought would be helpful, as well as similar to existing implementations in game-making software. Additionally, the authors of this paper incorporated multimedia learning principles wherever possible in developing the help facilities. Finally, a questionnaire was administered to the developers to verify that our objectives of consistent quality and realistic implementations was satisfied.

\subsection{Conditions}

We created 6 help facility conditions:

\begin{itemize}
	\item No Help
	\item Text Help
	\item Interactive Help
	\item Intelligent Agent Help
	\item Video Help
	\item All Help
\end{itemize}

\begin{figure*}[ht!] 
  \begin{minipage}[t]{0.5\linewidth}
    \centering
    \includegraphics[width=.8\linewidth]{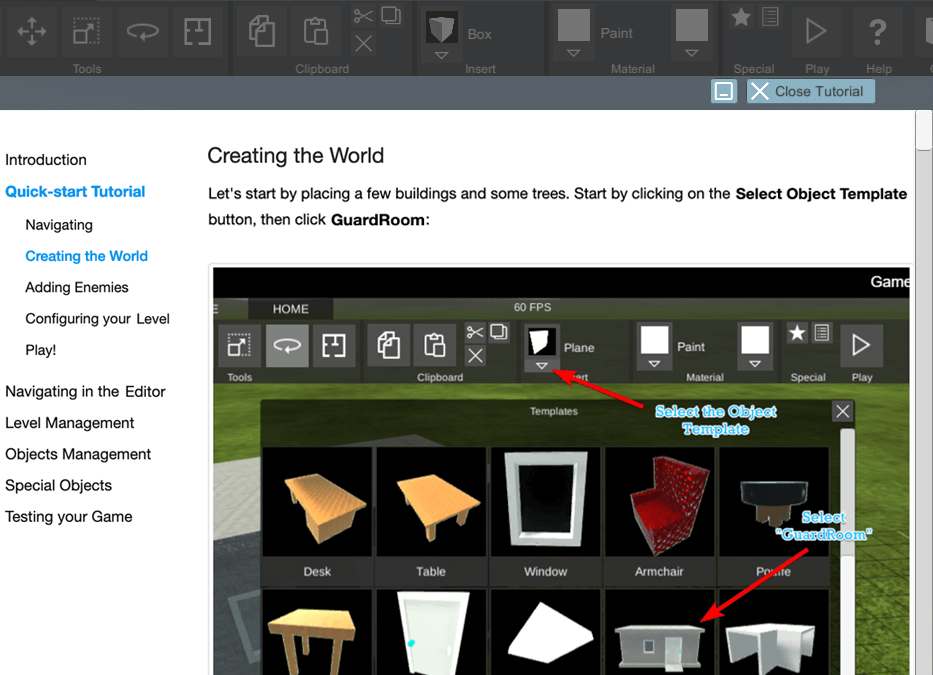} 
\caption{Text Help condition.}%
\Description{Fig6 description}
\label{fig:Condition1}
    \vspace{10px}
  \end{minipage}%
  \begin{minipage}[t]{0.5\linewidth}
    \centering
    \includegraphics[width=.8\linewidth]{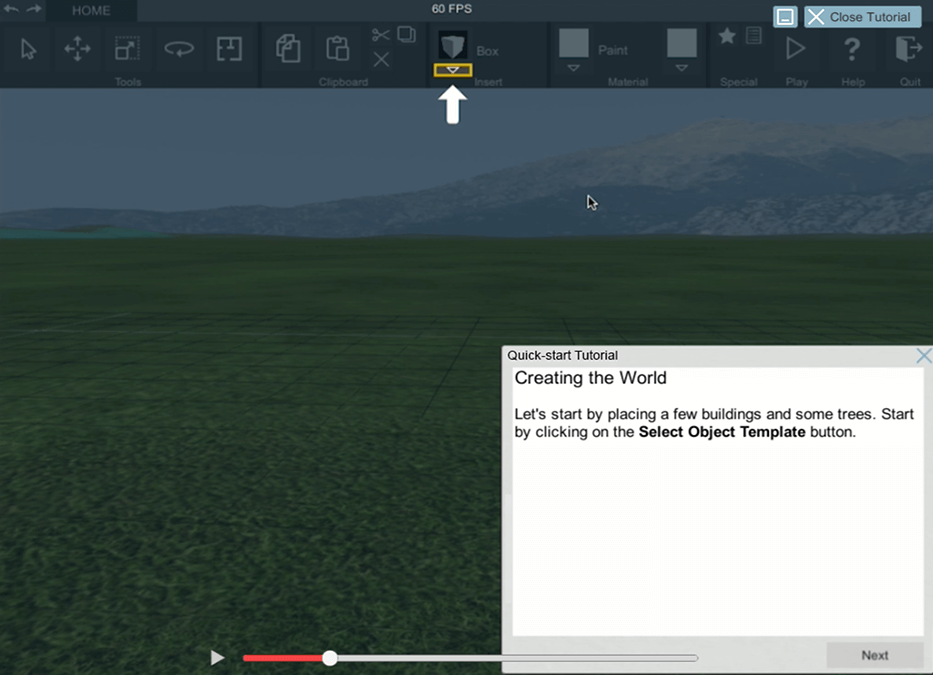} 
\caption{Video Help condition.}%
\Description{Fig7 description}
\label{fig:Condition4}
    \vspace{10px}
  \end{minipage} 
  \begin{minipage}[t]{0.5\linewidth}
    \centering
    \includegraphics[width=.8\linewidth]{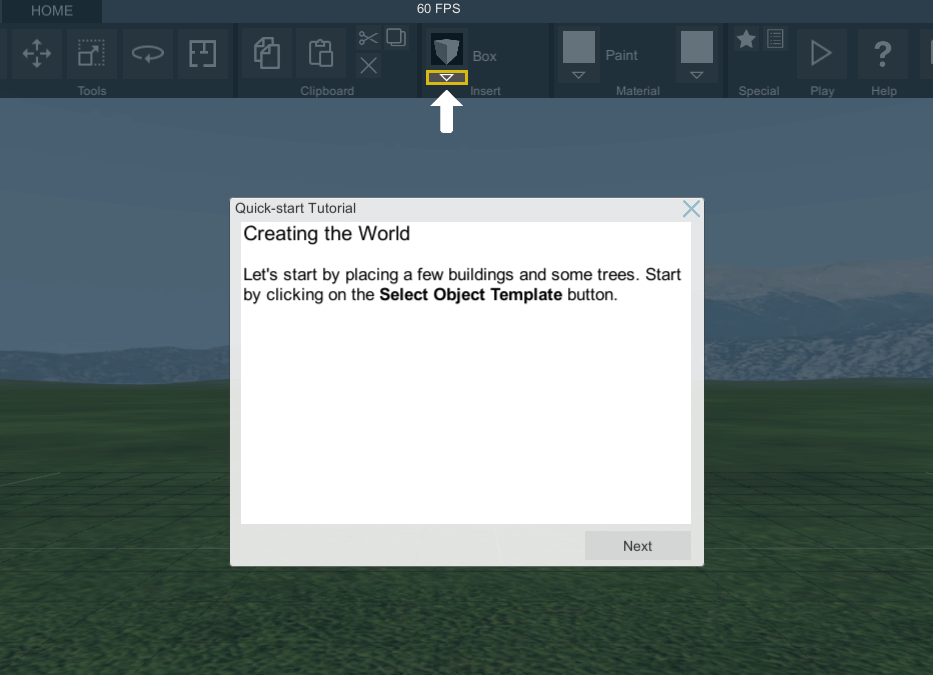} 
\caption{Interactive Help condition.}
\Description{Fig8 description}
\label{fig:Condition2}
  \end{minipage}%
  \begin{minipage}[t]{0.5\linewidth}
    \centering
    \includegraphics[width=.8\linewidth]{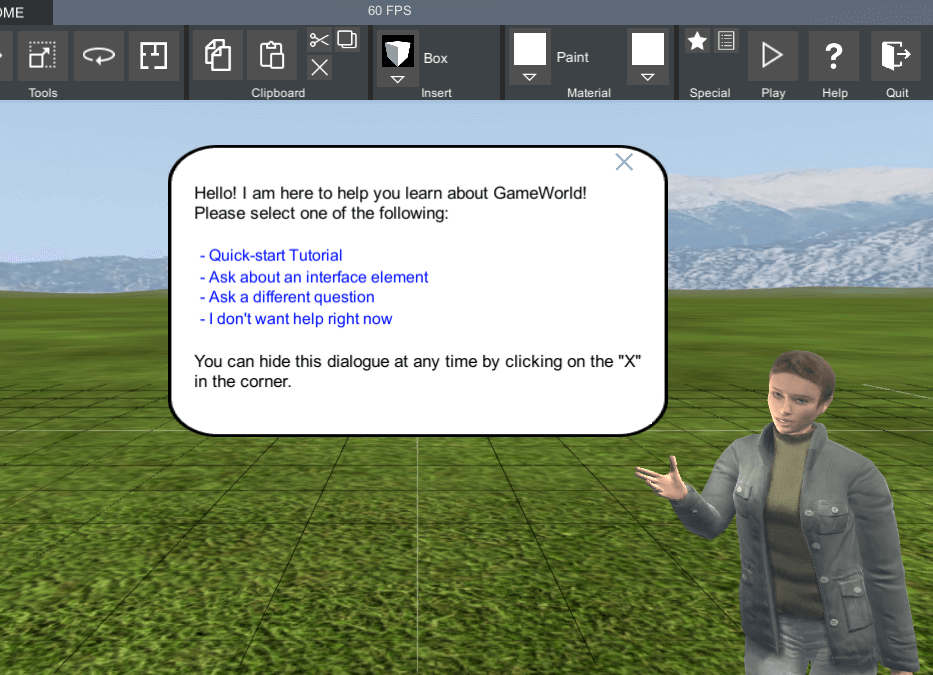} 
\caption{Intelligent Agent Help condition.}
\Description{Fig9 description}
\label{fig:Condition3}
  \end{minipage} 
\end{figure*}

See Figures~\ref{fig:Condition1},~\ref{fig:Condition4},~\ref{fig:Condition2}, and~\ref{fig:Condition3}. The No Help condition was a baseline control condition. The All Help condition contained all of the help facilities. Every help facility contained the same identical quick tutorial which consisted of learning \textit{navigation} (moving around the editor), \textit{creating the world} (creating and rotating objects), \textit{adding enemies} (creating enemy respawn points), \textit{level configuration} (changing additional level parameters), and \textit{play} (play testing). Upon completing the quick tutorial, users will have learned all of the concepts necessary to create their own level. Help facilities are integrated directly into the application to facilitate data tracking.

When the editor first loads, the user is presented with the dialog ``Go to X now?'' (X is replaced by Text Help, Interactive Help, Intelligent Agent Help, or Video Help). If the user clicks ``Yes'', the help facility is opened. If the user presses ``No'' then the user is notified that they can access the help facility at any time by clicking on the help icon. This happens only once for each of the help facility conditions. In the All Help condition, every help facility is presented in the opening dialog in a randomized order (randomized per-user), with each help facility presented as a button and ``No Thanks'' at the bottom of the list. In the No Help condition, no opening dialog is presented, and pressing the help icon brings up the dialog ``Currently Unavailable''.

\subsubsection{Text Help}

The Text Help window is a document that contains a menu bar and links to quickly navigate to different sections of the text help. The Text Help window can be closed or minimized. In either case, the Text Help window will re-open at the same location the user was at previously. When the Text Help is minimized, it appears as an icon near the bottom of the editor screen with the caption ``Text Help''.

The Text Help contains the quick tutorial. However, after the quick tutorial, the Text Help contains additional reference material. This includes in-depth (advanced) documentation on navigation, level management, objects management, special objects, and play testing. Additional information is provided that is not covered in the quick tutorial (e.g., hold down shift to amplify navigation, how to peek from behind corners during play, how to add lighting, etc.). Screenshots are provided throughout to add clarity.

\subsubsection{Interactive Help}

The Interactive Help provides the quick tutorial interactively. Players are limited to performing a specific action at each step (a darkened overlay only registers clicks within a cut-out area). When users are presented with information that does not require a specific action, users can immediately click ``Next''. Users can close the interactive tutorial at any time. If a user re-opens a previously closed interactive tutorial, the tutorial starts at the beginning---this behavior is consistent with existing interactive tutorials in game-making software.

\subsubsection{Intelligent Agent Help}

The Intelligent Agent Help is an intelligent agent that speaks to the user through dialog lines. A female voice actor provided the dialog lines of the intelligent agent. The intelligent agent has gestures, facial expressions, and lip movements that are synchronized to the audio. This was  facilitated using the \textit{SALSA With RandomEyes} and \textit{Amplitude for WebGL} Unity packages. For gestures, we created custom talk and idle animations.

When the Intelligent Agent Help is activated, it provides several options: 1) Quick Tutorial (identical to the interactive tutorial and everything is spoken by the intelligent agent), 2) Interface Questions (clicking anywhere on the screen provides an explanation---this also works with dialogs that are open such as level properties), 3) Other Questions (a pre-populated list of questions weighted by the user's least taken actions, e.g., if the user has already accessed level properties, this particular question will appear on a later page; there are three pages of common questions, e.g., ``How do I add enemies?''). The agent can be closed and re-activated at any time.

\subsubsection{Video Help}

The Video Help provides the quick tutorial in a video format (4 minutes, 27 seconds). Audio is voiced by the same female actor as for the Intelligent Agent Help. Captions are provided at the beginning of each section (e.g., ``Navigating''). A scrubber at the bottom allows the user to navigate the video freely. The Video Help can be closed or minimized. In either case, the Video Help window will re-open at the same location the user was at previously. When the Video Help is minimized, it appears as an icon near the bottom of the editor screen with the caption ``Video Help''.

\subsubsection{All Help}

In the All Help condition, users have access to all help facilities. Help facilities work the same as described, except only one help facility can be active at a time.

\subsection{Validating Help Facilities}

The feedback of the professional game developers played an important role in the creation of our help facilities. For example, the initial prototype of the Text Help was a simple text document with images. However, developers commented that most game-making software would contain easy-to-navigate documentation. Therefore, we enhanced the Text Help with a menu bar that contained section links.

After completing the final versions of the help facilities, we asked the game developers to answer a short survey. Each game developer, on their own, explored each help facility in a randomized order for at least 30 minutes. After each help facility, game developers anonymously answered two questions: ``Overall, I felt that the quality of the X was excellent,'' and ``Overall, I felt that the X was similar to how I would expect it to be implemented in other game-making software,'' on a scale of 1:\textit{Strongly Disagree} to 7:\textit{Strongly Agree}.

A one-way ANOVA found no significant effect of help facility condition on game developer quality ratings at the p<.05 level [F(3,56) = 0.24, p = 0.87]. The average quality score for each help facility was M=6.0, SD=1.3 (Interactive Tutorial), M=5.9, SD=1.1 (Video Tutorial), M=6.1, SD=0.8 (Text Tutorial), M=5.9, SD=0.8 (Intelligent Agent). A one-way ANOVA found no significant effect of help facility condition on game developer similar implementation ratings at the p<.05 level [F(3,56) = 0.35, p = 0.79]. The average similarity scores for each help facility was M=6.3, SD=0.7 (Interactive Tutorial), M=6.1, SD=0.7 (Video Tutorial), M=6.3, SD=0.8 (Text Tutorial), M=6.2, SD=0.7 (Intelligent Agent).

\subsection{Validating Frame Rate}

To ensure the validity of the experiment, one of our initial goals was to normalize frames per second across the help facilities. A lower frames-per-second count while one of the help facilities was active would present a possible experiment confound. We wanted to ensure that, in particular, the Intelligent Agent which is a 3D model that moved with gestures and facial expressions in a WebGL application, did not create performance issues and a possible degradation of the experience.

For testing, we used a 2018 PC (Windows 10) and a 2012 Macbook Pro (MacOS High Sierra). The PC had an Intel Core i7-7700k CPU (4.20 GHz), an NVIDIA GeForce GTX 1070, and 16 GB of RAM. The Mac had an Intel Core i5 (2.5 GHz), an Intel HD Graphics 4000 GPU, and 6 GB of RAM. Both systems used Firefox Quantum 63.0 to run the Unity WebGL game and for performance profiling.

We produced a 1-minute performance profile for each machine and for each help facility. In the case of Text Help, Interactive Help, and Intelligent Agent Help, interactions occurred at a reading speed of 200 words per minute \cite{Ziefle1998}. We produced a performance profile for each machine and for each help facility. All help facilities were within \textasciitilde1 fps: intelligent agent (PC: 59.14 fps, Mac: 59.04), interactive tutorial (PC: 60.00, Mac: 59.11), text tutorial (PC: 60.00, Mac: 59.43), video tutorial (PC: 60.00, Mac: 58.74).%

\section{Methods}

\subsection{Quantitative Measures}

\subsubsection{Learnability of Controls}

The ``Controls'' subscale from the Player Experience of Need Satisfaction (PENS) scale \cite{Ryan2006} was adapted for use in this study. This consisted of 3 questionnaire items as follows: ``Learning GameWorld's controls was easy'', ``GameWorld's controls are intuitive'', and ``When I wanted to do something in GameWorld, it was easy to remember the corresponding control''. Cronbach's alpha was 0.86.

\subsubsection{Learning Motivation Scale}

Learning motivation was captured using a scale adapted from \cite{hwang2011formative} which consisted of 7 items on a 6-point Likert scale (1: \textit{Strongly Disagree} to 6: \textit{Strongly Agree}), e.g., ``I would like to learn more about GameWorld''. Cronbach's alpha was 0.93.

\subsubsection{Cognitive Load}

Cognitive load used measures adapted from \cite{paas1992training} and \cite{sweller1998cognitive}. It consists of 8 items on a 6-point Likert scale (1: \textit{Strongly Disagree} to 6: \textit{Strongly Agree}). There are two sub-scales: mental load (e.g., ``GameWorld was difficult to learn for me''), and mental effort (e.g., ``Learning how to use GameWorld took a lot of mental effort''). Cronbach's alpha was 0.90 and 0.85.

\subsubsection{Game Quality Ratings}

Users were asked to rate their final game level on the dimensions of: ``Aesthetic'' (Is it visually appealing?), ``Originality'' (Is it creative?), ``Fun'' (Is it fun to play?), ``Difficulty'' (Is it difficult to play?), and ``Overall'' (Is it excellent overall?) on a scale of 1: \textit{Strongly Disagree} to 7: \textit{Strongly Agree}.

Expert ratings were given by 3 QA testers we hired. All QA testers had extensive games QA experience. The 3 QA testers first underwent one-on-one training with a GameWorld expert for one hour. QA testers then reviewed 250 game levels on their own without scoring them. QA testers were then given 50 game levels at random to rate. The GameWorld expert provided feedback on the ratings, and the game levels were rescored as necessary. Afterwards, QA testers worked entirely independently.

All 3 QA testers were blind to the experiment---the only information they received was a spreadsheet containing links to each participant's game level. Each game level was played by the QA tester before being rated. They were debriefed on the purpose of their work after they completed all 1646 ratings. The 3 QA testers each spent an average of 64 hours (SD=9.3) over 3 weeks, at \$10 USD/hr.

\subsubsection{Total Time}

We measure both total time, and time spent in each help facility. For all types of help, this is the amount of time that the help is on-screen (and maximized if it is Text Help or Video Help).

\subsubsection{Other Measures}

We were additionally interested in whether the player activated the help immediately on startup and how many total game-making actions were performed (this was an aggregate measure that combined object creations, object manipulations, etc.).

\subsection{Participants}

After a screening process that disqualified participants with multiple surveys with zero variance, multiple surveys with $\pm$3SD, or a failed attention check, 1646 Amazon Mechanical Turk participants were retained. The data set consisted of 976 male, and 670 female participants. Participants were between the ages of 18 and 73 (M = 32.3, SD = 9.6), were all from the United States, and could all read/write English. %

\subsection{Design}

A between-subjects design was used: help facility condition was the between-subject factor. Participants were randomly assigned to a condition.

\subsection{Protocol}

Participants filled out a pre-survey assessing previous experience playing games, programming, and creating games (conditions did not differ significantly across any of these measures, p=0.320, p=0.676, p=0.532). Then for a minimum of 10 minutes, each participant interacted with \textit{GameWorld}. After the 10 minutes had passed, the quit button became active and participants could exit at any time. After quitting, participants completed the PENS, the learning motivation scale, and the cognitive load scale. Participants then provided ratings on their game levels before filling out demographics.

\subsection{Analysis}

Separate MANOVAs are run for each separate set of items---\textit{PENS, User Level Ratings, Expert Level Ratings}; with the independent variable---\textit{help facility condition}. To detect the significant differences between badge conditions, we utilized one-way MANOVA. These results are reported as significant when p<0.05 (two-tailed). Prior to running our MANOVAs, we checked both assumption of homogeneity of variance and homogeneity of covariance by the test of Levene's Test of Equality of Error Variances and Box's Test of Equality of Covariance Matrices; and both assumptions were met by the data. For individual measures, we use one-way ANOVA.

\section{Results}

\begin{figure*}[ht!] 
\begin{minipage}[t]{0.24\linewidth}
\centering
\includegraphics[width=1\linewidth]{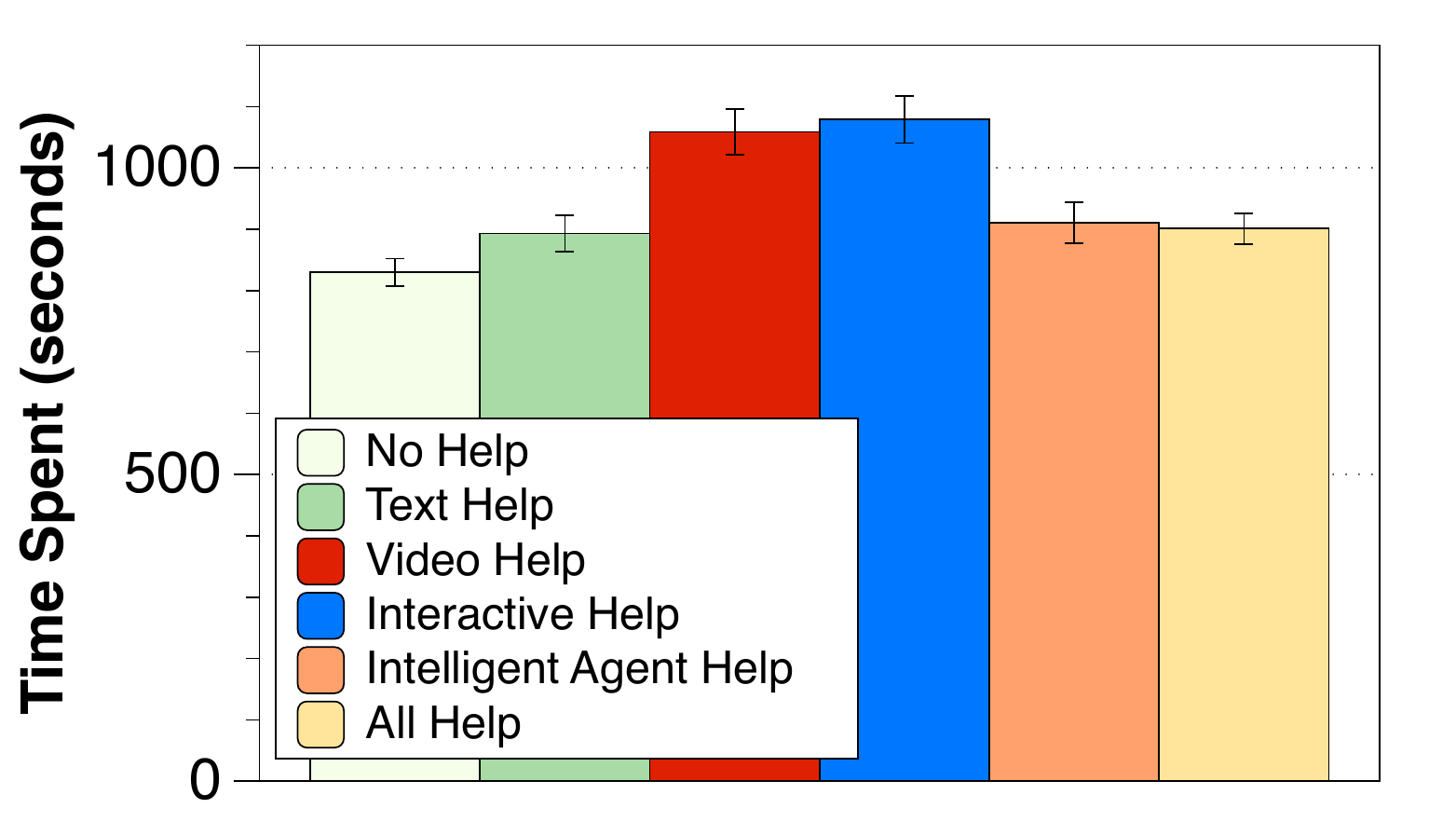}
\caption{Time spent  \mbox{(+/- SEM)}.}
\Description{Fig10 description}
\label{fig:TimeSpent}
\end{minipage}%
\hfill
\begin{minipage}[t]{0.24\linewidth}
\centering
\includegraphics[width=1\linewidth]{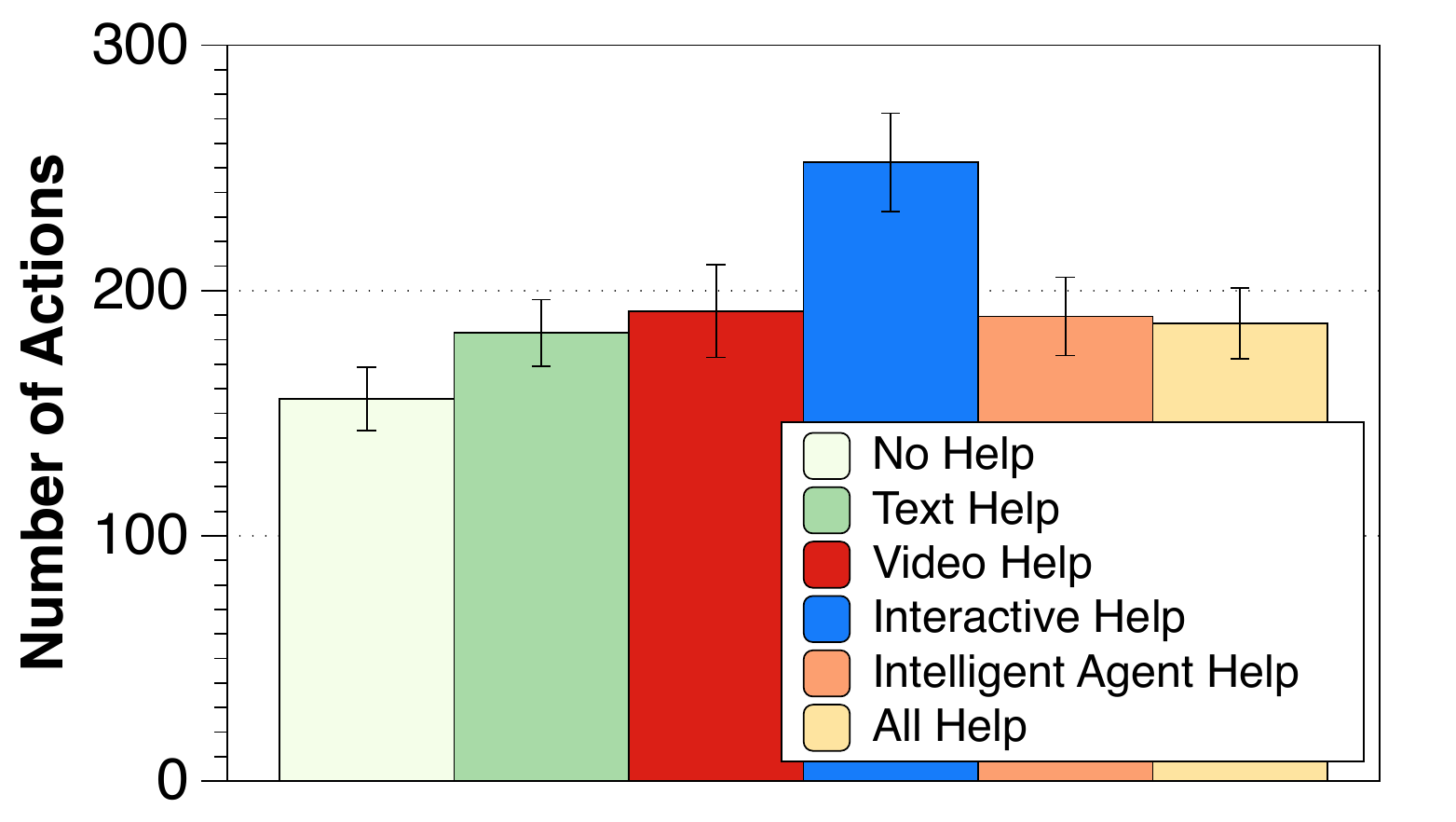}
\caption{Editor actions \mbox{(+/- SEM)}.}
\Description{Fig11 description}
\label{fig:TotalActions}
\end{minipage}%
\hfill
\begin{minipage}[t]{0.24\linewidth}
\centering
\includegraphics[width=1\linewidth]{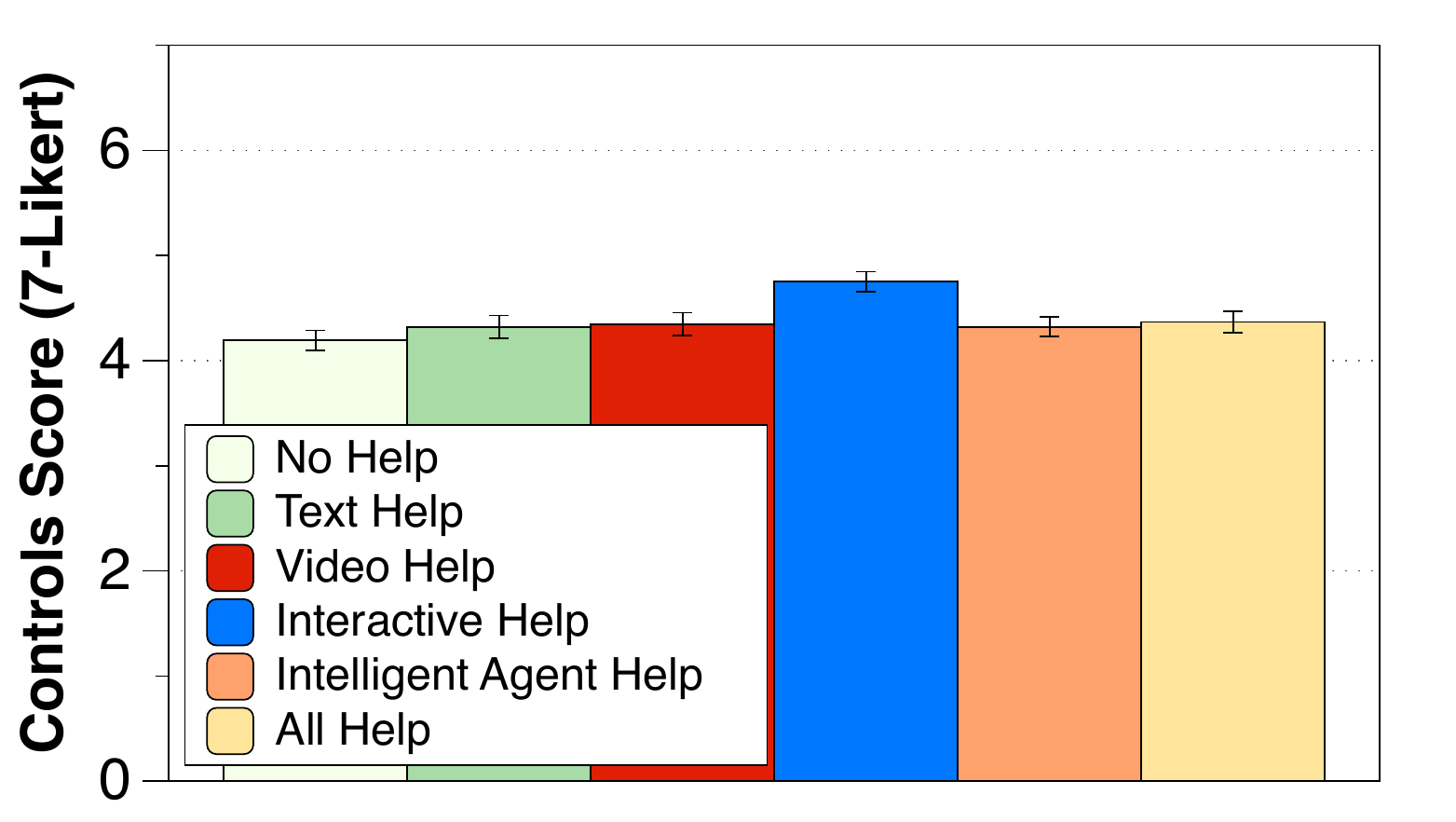}
\caption{Controls score \mbox{(+/- SEM)}.}
\Description{Fig12 description}
\label{fig:PENS}
\end{minipage}%
\hfill
\begin{minipage}[t]{0.24\linewidth}
\centering
\includegraphics[width=1\linewidth]{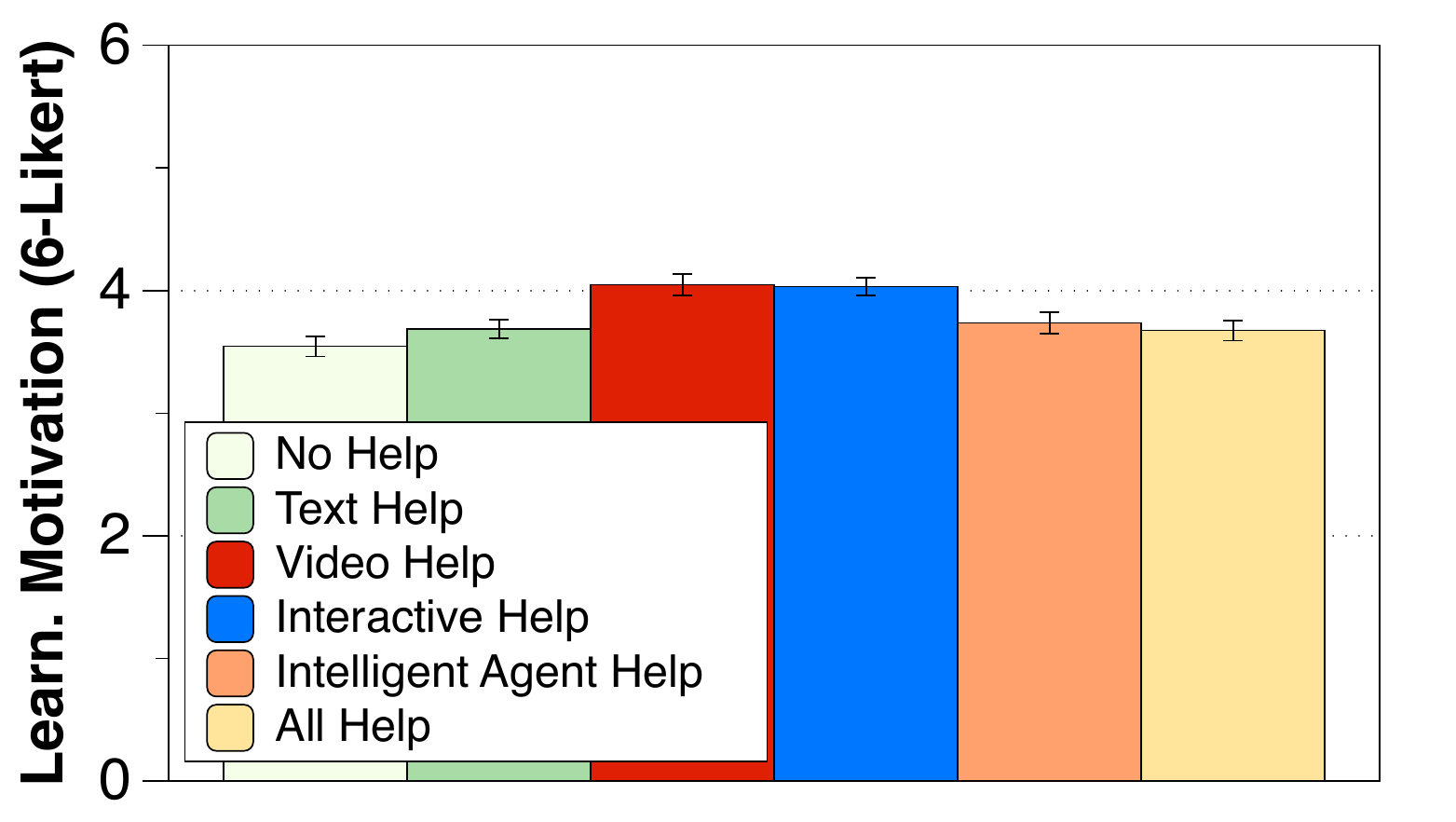}
\caption{Learning mot.  \mbox{(+/- SEM)}.}
\Description{Fig13 description}
\label{fig:LearningMotivation}
\end{minipage} 
\end{figure*}

\textbf{RQ1: Do help facilities lead to higher motivated behavior?} %

\textit{The Interactive Help and Video Help promoted greater time spent. The Interactive Help promoted a higher number of actions. The No Help condition results in the least time spent and the lowest number of actions.}

A one-way ANOVA found a significant effect of help facility condition on time spent at the p<.05 level [F(5,1640) = 10.23, p < 0.001, $\eta_{p}^{2}$ = 0.03]. Post-hoc testing using Tukey HSD found that participants in both the Interactive Help and Video Help conditions spent a longer total time than participants in any of the four other conditions, p<.05, \textit{d} in the range of 0.27--0.46. See Figure~\ref{fig:TimeSpent}.

A one-way ANOVA found a significant effect of help facility condition on total game-making actions at the p<.05 level [F(5,1640) = 4.22, p < 0.001, $\eta_{p}^{2}$ = 0.01]. Post-hoc testing using Tukey HSD found that participants in the Interactive Help condition performed a higher number of actions than  Text Help (\textit{d}=0.23), Intelligent Agent Help (\textit{d}=0.21), All Help (\textit{d}=0.22), and No Help (\textit{d}=0.33), p<.05. See Figure~\ref{fig:TotalActions}.

\noindent\textbf{RQ2: Do help facilities improve learnability of controls?}

\textit{The Interactive Help promoted controls learnability.}

A one-way ANOVA found a significant effect of help facility condition on the PENS controls score at the p<.05 level [F(5,1640) = 3.96, p < 0.005, $\eta_{p}^{2}$ = 0.01]. Post-hoc testing using Tukey HSD found that participants in  the Interactive Help condition had a higher PENS controls score than participants in any of the other conditions except All Help and Video Help, p \textless .05, \textit{d} in the range of 0.27--0.34. See Figure~\ref{fig:PENS}.

\noindent\textbf{RQ3: Do help facilities improve learning motivation?}

\textit{The Interactive Help and Video Help promoted learning motivation.  No Help results in the lowest learning motivation.}

A one-way ANOVA found a significant effect of help facility condition on learning motivation at the p<.05 level [F(5,1640) = 6.42, p < 0.001, $\eta_{p}^{2}$ = 0.02]. Post-hoc testing using Tukey HSD found that participants in both the Interactive Help and Video Help conditions had higher learning motivation than participants in any of the other conditions except Intelligent Agent Help, p<.05, \textit{d} in the range of 0.27--0.37. See Figure~\ref{fig:LearningMotivation}.

\begin{figure*}[ht!] 
\begin{minipage}[t]{0.49\linewidth}
\centering
\includegraphics[width=1\linewidth]{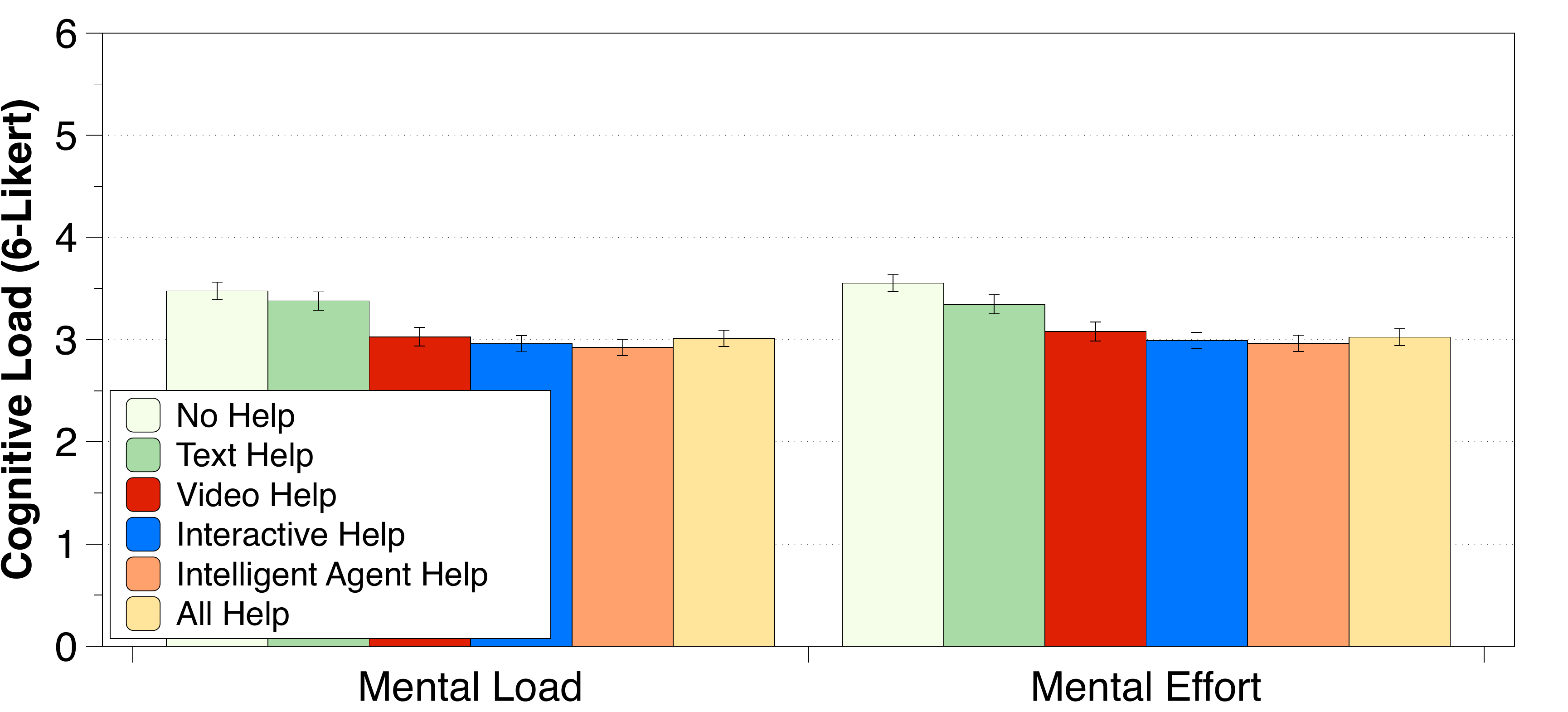}
\caption{Means of 6-point Likert ratings for cognitive load (+/- SEM).}
\Description{Fig14 description}
    \vspace{-10px}
\label{fig:CognitiveLoad}
\end{minipage}%
\hfill
\begin{minipage}[t]{0.49\linewidth}
\centering
\includegraphics[width=1\linewidth]{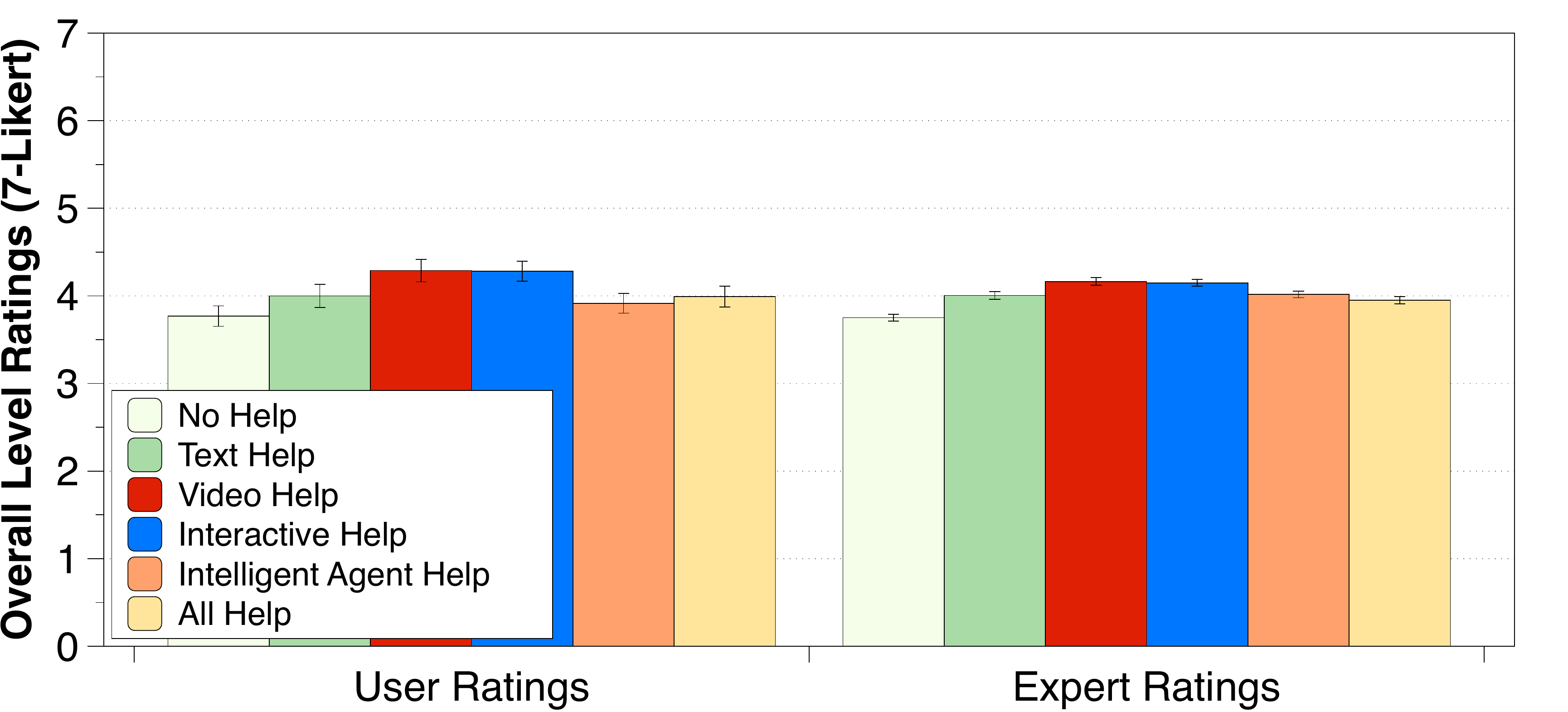}
\caption{Mean 7-point Likert \textit{overall} ratings for game levels (+/- SEM).}
\Description{Fig15 description}
    \vspace{-10px}
\label{fig:MapRatings}
\end{minipage}%
\end{figure*}

\noindent\textbf{RQ4: Do help facilities improve cognitive load?}

\textit{All conditions had lower cognitive load relative to No Help, except Text Help. Interactive Help, Intelligent Agent Help, and All Help have lower cognitive load than Text Help.}

A one-way ANOVA found a significant effect of help facility condition on mental load at the p<.05 level [F(5,1640) = 8.14, p < 0.001, $\eta_{p}^{2}$ = 0.02]. Post-hoc testing using Tukey HSD found that participants in the No Help condition had a higher mental load than participants in any other condition except Text Help, p<.005, \textit{d} in the range of 0.32--0.39. Participants in the Text Help condition had a higher mental load than Interactive Help (\textit{d}=0.31), Intelligent Agent Help (\textit{d}=0.33), and All Help (\textit{d}=0.28), p \textless .05.

A one-way ANOVA found a significant effect of help facility condition on mental effort at the p<.05 level [F(5,1640) = 8.29, p < 0.001, $\eta_{p}^{2}$ = 0.03]. Post-hoc testing using Tukey HSD found that participants in the No Help condition exerted higher mental effort than participants in any other condition, p<.005, \textit{d} in the range of 0.15--0.42. Participants in the Text Help condition exerted higher mental effort than Interactive Help (\textit{d}=0.26) and Intelligent Agent Help (\textit{d}=0.28), p \textless .05. See Figure~\ref{fig:CognitiveLoad}.

\noindent\textbf{RQ5: Do help facilities improve created game levels?}

\textit{The Interactive Help and Video Help led to the highest quality game levels, both from the user's perspective and expert ratings. No Help leads to the lowest quality.}

The MANOVA was statistically significant across help facility conditions across the self-rated game level quality dimensions, F(25, 6079) = 1.53, p \textless .05; Wilk's $\lambda$ = 0.977, $\eta_{p}^{2}$ = 0.01. ANOVAs found that the effect was significant across all dimensions except difficulty, p \textless .05, $\eta_{p}^{2}$ in the range of 0.01--0.02. Posthoc testing using Tukey HSD found that for aesthetic, originality, fun, and overall: Both Interactive Help and Video Help were significantly higher than No Help, p \textless .05, \textit{d} in the range of 0.25--0.39.

For expert ratings, intraclass correlation across the three raters was ICC=0.83 (two-way random, average measures), indicating high agreement. The MANOVA was statistically significant across help facility conditions across the expert-rated game level quality dimensions F(25, 6079) = 5.97, p \textless .001; Wilk's $\lambda$ = 0.914, $\eta_{p}^{2}$ = 0.02. ANOVAs found that the effect was significant across all dimensions, p \textless .005, $\eta_{p}^{2}$ in the range of 0.02--0.06. Posthoc testing using Tukey HSD found that Interactive Help and Video Help were highest across all dimensions (significant values: p \textless .05, \textit{d} in the range of 0.22--0.75). On the other hand, No Help was lowest across all dimensions (significant values: p \textless .05, \textit{d} in the range of 0.32--0.75). For the \textit{overall} dimension: Both Interactive Help and Video Help were significantly higher than all other conditions except Intelligent Agent Help and Text Help, p \textless .05, \textit{d} in the range of 0.26--0.58. No Help was lower than all other conditions, p \textless .05, \textit{d} in the range of 0.38--0.58. See Figure~\ref{fig:MapRatings}.

\noindent\textbf{RQ6: Does time spent on help facilities vary?}

\textit{The Interactive Help, Intelligent Agent Help, and Video Help lead to longest time spent on help. In the All Help condition, participants spent the most time in the Interactive Help, and next longest in Video Help. In the All Help condition, participants are less likely to activate any help facility on load.}

A one-way ANOVA found a significant effect of help facility condition on time spent on help at the p<.05 level [F(5,1640) = 36.85, p < 0.001, $\eta_{p}^{2}$ = 0.10]. Post-hoc testing using Tukey HSD found that participants in the Interactive Help (M=241, SD=313), Intelligent Agent Help (M=246, SD=382), and Video Help (M=238, SD=331), conditions spend more time on help than participants in Text Help (M=117, SD=198), and All Help (M=158, SD=202), p<.05, \textit{d} in the range of 0.29--0.42.

A one-way within subjects ANOVA was conducted to compare time spent across different help facilities in the All Help condition. There was a significant difference in time spent, Wilk's $\lambda$ = 0.929, F (3,275) = 7.04, p < .001, $\eta_{p}^{2}$ = 0.07. Post-hoc testing using a Bonferroni correction found that participants in the All Help condition spent significantly longer in the Interactive Help (M=63, SD=135) than in the Text Help (M=24, SD=81, \textit{d}=0.35) and the Intelligent Agent Help (M=24, SD=86, \textit{d}=0.34), p<.001. Time spent in Video Help was M=47, SD=144.

A one-way ANOVA found a significant effect of help facility condition on likelihood of startup help activation at the p<.05 level [F(5,1640) = 282.64, p < 0.001, $\eta_{p}^{2}$ = 0.46]. Post-hoc testing using Tukey HSD found that participants in the All Help condition (M=66\%) are significantly less likely to activate a help facility on startup than any other help facility condition, p<.05, \textit{d} in the range of 0.20--0.58.

\section{Discussion}

\subsection{Game Making Help Facilities Are Crucial}

The results show that Interactive Help promoted time spent, total editor activity, controls learnability, and learning motivation. Video Help promoted time spent, and learning motivation. These results highlight the important role that help facilities had in promoting motivated behavior, controls learnability, and learning motivation in \textit{GameWorld}.%

For cognitive load, No Help had the highest load with Text Help the second highest. All other conditions had lower cognitive load. These results show that help facilities can reduce the cognitive load for users, and that help facilities (e.g., Text Help) can have differential impacts.

Finally, results show that help facilities improve the quality of produced game levels. Both Interactive Help and Video Help led to the highest quality game levels, both self and expert rated. On the other hand, No Help led to the lowest quality. This demonstrates that help facilities improve the objective quality of produced artifacts in \textit{GameWorld}.

\subsection{Not All Help Facilities Are Made Equal}

Results show that No Help is detrimental to most outcomes. Having some help facility was better than having no help facility. However, there was significant variance between help facilities. Text Help only marginally improved outcomes compared to No Help. Similarly, All Help and Intelligent Agent Help saw only marginal improvements compared to No Help, with the exception of cognitive load (on which both All Help and Intelligent Agent Help scored low). On the other hand, the results show that Interactive Help and Video Help led to significant improvements over No Help with medium--small effect sizes \cite{cohen1992power}.

The results show that participants in the All Help condition are less likely to activate a help facility on startup. This potentially indicates too much choice, or a choice that was simply not meaningful \cite{Flowerday2000,Katz2007,Rose2002,Evans2015}. On the other hand, the two effective help facilities, Interactive Help and Video Help, are linear. These two help facilities are also the ones that participants in the All Help condition spent the most time with, suggesting that users preferred to spending time in these help facilities over Text Help and Intelligent Agent Help.

\subsection{Why These Findings Occurred}

Both Interactive Help and Video Help outperformed other conditions. Both Interactive Help and Video Help are able to moderate cognitive load in comparison to No Help and Text Help, through allowing users to follow guided step-by-step instructions. A reduction in cognitive load often results in better performance \cite{Paas1992}, which in this context translated to time spent, editor actions, controls learnability, learning motivation, and game level quality. The additional benefits of Interactive Help above and beyond other conditions in this study could be a result of performing actions immediately as they are being described during instruction. For example, decades of studies have shown the effectiveness of learning techniques that involve the act of doing while learning, including hands-on learning \cite{regan1996interactive}, active learning \cite{silberman1996active}, and situated learning \cite{Lave1991a}. In a meta-analysis of 255 studies, active learning---which promotes directly interacting with learning material \cite{Bonwell1991}---was shown to reduce failure rates in courses by 11\% and increase student performance on course assessments by 0.47 standard deviations \cite{silberman1996active}. Therefore, interactive help facilities have a strong theoretical and empirical basis for their effectiveness. More work, however, is needed to understand why Intelligent Agent Help was less helpful than Interactive Help. It is possible that the number of dialog choices contained in the Intelligent Agent Help was overwhelming for users \cite{Flowerday2000,Katz2007,Rose2002,Evans2015}. More research is needed to understand how to best optimize different help facilities.

\subsection{Recommendations for Game Making Software}

\textbf{Interactive Help and Video Help Improved Outcomes.} Our results show that Interactive Help and Video Help lead to improved outcomes. However, in our review of game-making software, we found that while 89.4\% had text documentation, only 52.9\% had videos and 20.0\% interactive tutorials. This indicates a potential missed opportunity for game-making software to better introduce systems to users\footnote{One aspect not analyzed in this study is cost/ease of development, which may be a reason for Text Help's ubiquity.}.

\noindent\textbf{Any Help Is Better Than No Help.} Participants in No Help performed badly on all outcomes (time spent, controls learnability, learning motivation, cognitive load, total editor activity, and game level quality). Having some help facility was always better than having no help facility at all. This indicates that game-making software should always incorporate some form of help, even if simply basic documentation.

\noindent\textbf{Be Wary of Giving Users Choice of Help.} Results show that participants in the All Help condition, in which participants were able to choose which help facility to use, led to worse outcomes than Interactive Help or Video Help alone. Participants in All Help were less likely to activate help on startup than any other condition, and spent less time on help compared to Interactive Help, Video Help, and Intelligent Agent Help. This indicates that initially prompting the user with one good help facility will be more effective.

\section{Limitations}

Amazon Mechanical Turk (AMT) has been shown to be a reliable platform for experiments (e.g., \cite{Mason2012,buhrmester2011amazon}). AMT workers also tend to represent a more diverse sample than the U.S. population~\cite{chandler2016conducting,buhrmester2011amazon,paolacci2010running}. However, future experiments restricted to experienced game developers could give more insight into help facilities' effects on experts. Despite that our AMT sample consists mainly of novices, these are likely the users who need the most scaffolding, and hence are an appropriate population to study in the context of help.

Longitudinal studies are needed. Although in the short-term, our results show that Interactive Help and Video Help are highly beneficial, the other help facilities could become more effective over time. For example, long-time users may find Text Help useful for looking up reference information. Longitudinal studies may determine, for example, that certain types of help are more appropriate for different levels of experience.

We took care to design help facilities in consultation with highly experienced game developers. Moreover, we ensured that game developers perceived the help facilities to be of high/similar quality, and that the help facilities were implemented similarly to other game-making software. Nevertheless, these help facilities could be constructed differently. For example, the intelligent agent could have been constructed to be similar to the user. A significant literature has shown that intelligent agents that are more similar to their users (termed similarity-attraction \cite{Byrne1965,Isbister2000}) along the dimensions of age, gender, race, clothing, etc. promote learning \cite{Kim2006,Baylor2004,Guadagno2007,Pratt2007,Rosenberg-Kima2010,Johnson2013,Arroyo2009,Bailenson2008b}. Indeed, there are any number of changes to these help facilities that we can imagine. Nonetheless, there is value in contrasting the baseline developer-driven and developer-validated implementations here.

Finally, there are other forms of help that we are interested in. For example, providing template projects can be useful both as a starting point and for dissecting/understanding pre-built games. Additionally, we are interested in augmenting \textit{GameWorld} with additional game genres (e.g., platformer, action RPG, etc.), and capabilities.

\vspace{-2px}
\section{Conclusion}

Game-making is increasingly pervasive, with an ever-larger number of game engines and software. With today's game-making software, it is increasingly becoming possible for novices and experts alike to create games. Nonetheless, game-making software is often complex. Help facilities can therefore play an important role in scaffolding knowledge in game-making. Results show that Interactive Help was the most promising help facility, leading to a greater positive impact on time spent, controls learnability, learning motivation, total editor activity, and game level quality. Video Help is a close second across these same measures. These results are directly relevant to designers, researchers, and developers, as they reveal how to best support novice game-making through help facilities. Future research in this domain can help cultivate the next generation of game-makers in an age where play and making are, more than ever, both ubiquitous and intertwined.
\bibliographystyle{ACM-Reference-Format}
\bibliography{bib}


\end{document}